\documentclass[10pt,final,letter,twoside]{IEEEtran}

\usepackage{cite}
\usepackage{amsmath}
\usepackage{array}
\usepackage{graphicx}
\usepackage{url}

\newcommand{\argmin}{\operatornamewithlimits{argmin}}
\newcommand{\HT}{\mathsf{H}}

\usepackage{color}

\newcommand{\WPE}{WPE dereverberation}
\newcommand{\MPDR}{MPDR beamforming}
\newcommand{\MVDR}{MVDR beamforming}
\newcommand{\WPD}{WPD beamforming}

\begin{document}

\title{A unified convolutional beamformer for simultaneous denoising and dereverberation}

\author{Tomohiro~Nakatani,~\IEEEmembership{Senior Member,~IEEE,}
        Keisuke~Kinoshita,~\IEEEmembership{Senior Member,~IEEE}%
\thanks{T. Nakatani and K. Kinoshita are with NTT Communication Science Laboratories, NTT Corporation.}%
\thanks{Manuscript received December 19, 2018.}
}

\markboth{Copyright (c) 2019 IEEE. This is the author's version. The final version is at http://dx.doi.org/10.1109/LSP.2019.2911179}
{Copyright (c) 2019 IEEE. This is the author's version. The final version is at http://dx.doi.org/10.1109/LSP.2019.2911179}

\maketitle

\begin{abstract}
This paper proposes a method for estimating a convolutional beamformer that can perform denoising and dereverberation simultaneously in an optimal way. The application of dereverberation based on a weighted prediction error (WPE) method followed by denoising based on a minimum variance distortionless response (MVDR) beamformer has conventionally been considered a promising approach, however, the optimality of this approach cannot be guaranteed. To realize the optimal integration of denoising and dereverberation, we present a method that unifies the WPE dereverberation method and a variant of the MVDR beamformer, namely a minimum power distortionless response (MPDR) beamformer, into a single convolutional beamformer, and we optimize it based on a single unified optimization criterion. The proposed beamformer is referred to as a Weighted Power minimization Distortionless response (WPD) beamformer. Experiments show that the proposed method substantially improves the speech enhancement performance in terms of both objective speech enhancement measures and automatic speech recognition (ASR) performance.
\end{abstract}


\IEEEpeerreviewmaketitle

\section{Introduction}
When a speech signal is captured by distant microphones, e.g., in a conference room, it will inevitably contain additive noise and reverberation components. These components are detrimental to the perceived quality of the observed speech signal and often cause serious degradation in many applications such as hands-free teleconferencing and automatic speech recognition (ASR).

Microphone array signal processing techniques have been developed to minimize the aforementioned detrimental effects by reducing the noise and the reverberation in the acquired signal.  A filter-and-sum beamformer \cite{Anguera07ASLP}, a minimum-variance distortionless response (MVDR) beamformer and a minimum-power distortionless response (MPDR) beamformer \cite{MPDR,Cox,cgmmmvdr,erdogan16intersp,L1MPDR}, and a maximum signal-to-noise ratio beamformer \cite{GEV,maxSNR,BeamNET} are widely-used systems for denoising, while a weighted prediction error (WPE) method and its variants \cite{wpe,gwpe,owpe,swpe,cwpe} are emerging techniques for dereverberation. The usefulness of these techniques, particularly for improving ASR performance, has been extensively studied, e.g., at the REVERB challenge \cite{REVERB} and the CHiME-3/4/5 challenges \cite{CHiME3,CHiME4,CHiME5}. Advances in this technological area have led to recent progress on commercial devices with far-field ASR capability, such as smart speakers \cite{GoogleHome,homepod,SPM}.

However, it remains a challenge to reduce both noise and reverberation simultaneously in an optimal way. For example, researchers have proposed using {\MVDR} and {\WPE} in a cascade manner \cite{delcroix15eurasip,mvdrwpe}, where, for example, the signal is first processed by {\WPE} and then denoised with {\MVDR}.  With this approach, dereverberation may not be optimal due to the influence of the noise, and denoising may be disturbed by the remaining reverberation.  Certain joint optimization techniques have also been proposed \cite{Togami,lukas2018interspeech,wpegsc}, but they perform dereverberation and denoising separately, which makes the optimality of the integration unclear, resulting in marginal performance improvement compared with the cascade system.

To achieve optimal integration, this paper proposes a method for unifying {\WPE} and {\MPDR}, into a single convolutional beamforming approach and for optimizing the beamformer based on a single unified optimization criterion.   We can derive a closed-form solution for this beamformer, assuming that the time-varying power and steering vector of the desired signal are given.  The optimality of the beamformer is guaranteed under the assumed optimization criterion and condition.
The beamformer is referred to as a Weighted Power minimization Distortionless response (WPD) beamformer.
%
Note that the steering vector and the signal power must also be given for {\WPE} and {\MPDR}, respectively, and several techniques for their estimation have already been proposed \cite{lukas2018interspeech,ito17icassp,Gannot15icassp}.

In the experiments, we compare the proposed method with {\WPE}, {\MPDR}, and both approaches in a cascade configuration in terms of objective speech enhancement measures and ASR performance.  The experiments show that the proposed method substantially outperforms all the conventional methods with regard to almost all the performance metrics.  For example, in comparison with the cascade system, the proposed method achieves an average word error reduction rate of 7.5 \% for real data taken from the REVERB Challenge dataset.


\section{Signal model}
Assume that a single speech signal is captured by $M$ microphones in a noisy reverberant environment. Then, the captured signal in the short time Fourier transform (STFT) domain is approximately modeled at each frequency bin by 
\begin{equation}
\mathbf{x}_{t}=\sum_{\tau^=0}^{L_a} \mathbf{a}_{\tau}s_{t-\tau}+\mathbf{n}_{t},
\label{eq:obs}
\end{equation}
where $t$ and $\tau$ are time frame indices. 
Note that all the symbols should also have frequency bin indices, but they are omitted for brevity in this paper assuming that each frequency bin is processed independently in the same way. 
Letting $\top$ denote the non-conjugate transpose, $\mathbf{x}_{t}=[x_{t}^{(1)},x_{t}^{(2)},\ldots,x_{t}^{(M)}]^{\top}$ is a column vector containing the STFT coefficients of the captured signals for all the microphones at a time frame $t$, $s_{t}$ is an STFT coefficient of clean speech signal at a time frame $t$, $\mathbf{a}_{t}=[a_{t}^{(1)},a_{t}^{(2)},\ldots,a_{t}^{(M)}]^{\top}$ for $t=0,1,\ldots,L$ is a sequence of column vectors containing convolutional acoustic transfer functions (ATFs) from the speaker location to all the microphones, $L_a$ is the length of the convolutional ATFs in each frequency bin, and $\mathbf{n}_{t}=[n_{t}^{(1)},n_{t}^{(2)},\ldots,n_{t}^{(M)}]^{\top}$ is the additive noise.  As in eq.~(\ref{eq:obs}), according to \cite{Nak-ICASSP2008}, the effect of the reverberation can be approximately represented by the convolution in the STFT domain between $s_{t}$ and $\mathbf{a}_{t}$ when the length of the room impulse response in the time domain is longer than the analysis window. Hereafter, we refer to a sequence of STFT coefficients in each frequency bin, such as $x_{t}^{(m)}$ and $s_{t}$ for $t=1,2,\ldots$, simply as a signal.

The first term in eq.~(\ref{eq:obs}) can be further decomposed into two parts, one composed of a direct signal and early reflections, hereafter referred to as the desired signal $\mathbf{d}_{t}$, and the other corresponding to the late reverberation $\mathbf{r}_{t}$ \cite{early}. With this decomposition, eq.~(\ref{eq:obs}) is rewritten as
\begin{align}
\mathbf{x}_{t}&=\mathbf{d}_{t}+\mathbf{r}_{t}+\mathbf{n}_{t},\label{eq:obs2}\\
\mathbf{d}_{t}&=\sum_{\tau^=0}^{b-1} \mathbf{a}_{\tau}s_{t-\tau},\label{eq:desired}\\
\mathbf{r}_{t}&=\sum_{\tau^=b}^{L_a} \mathbf{a}_{\tau}s_{t-\tau},
\end{align}
where $b$ is the frame index that divides the convolutional ATFs into the ATF coefficients for $\mathbf{d}_{t}$ and those for $\mathbf{r}_{t}$. Later, $b$ is also termed the prediction delay for {\WPE} and {\WPD}. 
Finally, we define the goal of realizing speech enhancement to preserve $\mathbf{d}_{t}$ while reducing $\mathbf{r}_{t}$ and $\mathbf{n}_{t}$ from $\mathbf{x}_{t}$.

\section{Conventional methods}
This section gives a brief overview of the conventional methods, including {\WPE}, {\MPDR}, and two approaches with a cascade configuration.
\subsection{Dereverberation by WPE}
If we disregard the additive noise, $\mathbf{n}_{t}$, we can rewrite eq.~(\ref{eq:obs}) using a multichannel autoregressive model \cite{pe97,mstep,wpe} as
\begin{equation}
\mathbf{x}_{t}=\sum_{\tau=b}^{L_w}W_{\tau}^{\HT}\mathbf{x}_{t-\tau}+\mathbf{d}_{t},
\end{equation}
where $L_w$ is the regression order, $\HT$ denotes the conjugate transpose, $W_{t}$ for $t=b,b+1,\ldots,{L_w}$ are $M\times M$ dimensional matrices containing coefficients that predict the current captured signal, $\mathbf{x}_{t}$, from the past captured signals, $\mathbf{x}_{t-\tau}$ for $\tau=b,b+1,\ldots,{L_w}$,
and the second term in the equation, referred to as the prediction error, 
is assumed to be the desired signal according to the model \cite{wpe}.

{\WPE} estimates the prediction coefficients based on maximum likelihood estimation, assuming that the desired signal at each microphone follows a time-varying complex Gaussian distribution with a mean of zero and a time-varying variance, $\sigma_{t}^2$, which corresponds to the time-varying power of the desired signal. Then, the prediction coefficients, 
$\bar{W}=[W_{b},W_{b+1},\ldots,W_{L_w}]^{\top}$, 
are estimated as those that minimize the average power of the prediction error weighted by the inverse of $\sigma_{t}^2$. The estimation is represented by
\begin{equation}
\hat{\bar{W}}=\argmin_{\bar{W}}\sum_{t}\frac{\|\mathbf{x}_{t}-\sum_{\tau=b}^{L_w}W_{\tau}^{\HT}\mathbf{x}_{t-\tau}\|_2^2}{\sigma_{t}^{2}},
\label{eq:wpeest}
\end{equation}
where $||\mathbf{x}||_2^2=\mathbf{x}^{\HT}\mathbf{x}$ is the squared $L_2$ norm of a vector $\mathbf{x}$. It is known that the prediction delay $b$ also works as a distortionless constraint to prevent the desired signal components from being distorted by the dereverberation \cite{wpe}. As for the estimation of $\sigma_{t}^2$, several useful techniques have been proposed including an iterative estimation method \cite{Nak-ICASSP2008,swpe}. 

With the estimated prediction coefficients, the dereverberation is performed by
\begin{equation}
\hat{\mathbf{d}}_{t}=\mathbf{x}_{t}-\sum_{\tau=b}^{L_w}\hat{W}_{\tau}^{\HT}\mathbf{x}_{t-\tau}.
\label{eq:wpefilter}
\end{equation}
It was experimentally confirmed that {\WPE} can function robustly even in noisy environments to reduce the late reverberation with a slight increase in the noise \cite{wpe}.

\subsection{Beamforming by MPDR}
Assuming that 
the desired signal can be approximated as the product of a vector $\mathbf{v}$ with a clean speech signal, i.e., $\mathbf{d}_{t}=\mathbf{v}s_{t}$, and taking the late reverberation, $\mathbf{r}_{t}$, as part of the noise, $\mathbf{n}_{t}$, eq.~(\ref{eq:obs2}) becomes
\begin{equation}
  \mathbf{x}_{t}=\mathbf{v}s_{t}+\mathbf{n}_{t}.
\end{equation}
The MPDR beamformer is defined as a vector, $\mathbf{w}_{0}$, that minimizes the average power of the captured signal, $\mathbf{x}_{t}$, under a distortionless constraint, $\mathbf{w}_{0}^{\HT}\mathbf{v}=1$, that keeps the clean speech, $s_{t}$, unchanged by the beamforming \cite{MPDR,Cox}.  Here, $\mathbf{v}$ is also termed a steering vector, and techniques for its estimation from a captured signal have been proposed. Due to the scale ambiguity in the steering vector estimation, in practice it is substituted by a relative transfer function (RTF) \cite{RTF}.  An RTF is defined as the steering vector normalized by its value at a reference channel, calculated by $\mathbf{v}/v^{(q)}$ where $v^{(q)}$ denotes the value at the reference channel. 
This makes the distortionless constraint work to keep the desired signal at the reference channel, $d_{t}^{(q)}$, unchanged.

The beamformer is estimated as follows:
\begin{equation}
\hat{\mathbf{w}}_{0}=\argmin_{\mathbf{w}_{0}}\sum_{t}\left|\mathbf{w}_{0}^{\HT}\mathbf{x}_{t}\right|^2
~~\mathrm{s.t.}~~\mathbf{w}_{0}^{\HT}\mathbf{v}=1.
\label{eq:mpdrest}
\end{equation}
The desired signal is then estimated as
\begin{equation}
\hat{d}_{t}^{(q)}=\hat{\mathbf{w}}_{0}^{\HT}\mathbf{x}_{t}.
\label{eq:mpdrfilter}
\end{equation}
With the beamformer, the resultant signal is composed of only one channel signal corresponding to the reference channel $q$. 

On the basis of the above discussion, {\MPDR} can perform both denoising and dereverberation \cite{MPDRREVERB} by reducing $\mathbf{n}_{t}$, which contains the additive noise and the late reverberation. However, its dereverberation capability is limited because it cannot reduce reverberation components that come from the target speaker direction, especially when there are few microphones.

\subsection{Cascade of {\WPE} and {\MPDR}}
To achieve better speech enhancement in noisy reverberant environments, researchers have proposed using both {\WPE} and {\MPDR} in a cascade configuration \cite{delcroix15eurasip}.  Because {\WPE} can dereverberate all the microphone signals individually, {\MPDR} can be applied after {\WPE} has been applied. Techniques have also been proposed for estimating the steering vector and the power of the desired signal, for example, by iteratively and alternately applying {\WPE} and {\MPDR} to the signals \cite{lukas2018interspeech}.  

\section{Proposed method}
This section describes a method for unifying {\WPE} and {\MPDR} into a single convolutional beamforming approach. A closed-form solution can be obtained for the beamformer given the steering vector and the time-varying power of the desired signal, and we can perform more effective speech enhancement than with a simple cascade consisting of {\WPE} and {\MPDR}.
Figure~1 illustrates the processing flow of the method.

\subsection{Convolutional beamforming by WPD}
First, the signal obtained using the cascade consisting of {\WPE} and {\MPDR}, i.e., eqs.~(\ref{eq:wpefilter}) and (\ref{eq:mpdrfilter}), can be rewritten as
\begin{align}
\hat{d}_{t}^{(q)}&=\mathbf{w}_{0}^{\HT}\left(\mathbf{x}_{t}-\sum_{\tau=b}^{L_w}W_\tau^{\HT}\mathbf{x}_{t-\tau}\right),\\
&=\mathbf{w}_{0}^{\HT}\mathbf{x}_{t}+\sum_{\tau=b}^{L_w}\mathbf{w}_\tau^{\HT}\mathbf{x}_{t-\tau},
\label{eq:composite}\\
&=\bar{\mathbf{w}}^{\HT}\bar{\mathbf{x}}_{t},\label{eq:vform}
\end{align}
where we set $\mathbf{w}_{t}=-W_{t}\mathbf{w}_{0}$ to obtain the second line above, 
and we set $\bar{\mathbf{w}}=[\mathbf{w}_{0}^{\top},\mathbf{w}_{b}^{\top},\mathbf{w}_{b+1}^{\top},\ldots,\mathbf{w}_{L_w}^{\top}]^{\top}$ and $\bar{\mathbf{x}}_{t}=[\mathbf{x}_{t}^{\top},\mathbf{x}_{t-b}^{\top},\mathbf{x}_{t-b-1}^{\top},\ldots,\mathbf{x}_{t-L_w+1}^{\top}]^{\top}$ to obtain the third line. 
Note that $\bar{\mathbf{w}}$ and $\bar{\mathbf{x}}_{t}$ contain a time gap between their first and the second elements, corresponding to the prediction delay $b$. 

Next, the optimization criterion is defined based on the model of the desired speech used for {\WPE}, namely the time-varying Gaussian distribution, and based on the distortionless constraint used for {\MPDR}.  Specifically, we estimate the convolutional filter, $\bar{\mathbf{w}}$, as one that minimizes the average weighted power of a signal under a distortionless constraint.
It is represented by
\begin{equation}
  \hat{\bar{\mathbf{w}}}=\argmin_{\bar{\mathbf{w}}}\sum_{t}
  \frac{|\bar{\mathbf{w}}^{\HT}\bar{\mathbf{x}}_{t}|^2}{\sigma_{t}^{2}}
~\mathrm{s.t.}~\mathbf{w}_{0}^{\HT}\mathbf{v}=1.
\label{eq:wmpdrest}
\end{equation}
Here, all the filter coefficients are optimized based on the average weighted power minimization criterion. 
%
Note that the use of the time-varying weight makes the distribution of the enhanced speech obtained by beamforming closer to that of the desired speech.  

Eq.~(\ref{eq:wmpdrest}) can be viewed as a variation of eq.~(\ref{eq:mpdrest}), which is used for conventional {\MPDR}. Unlike eq.~(\ref{eq:mpdrest}), eq.~(\ref{eq:wmpdrest}) evaluates the average weighted power of the signal, and considers both the spatial and temporal covariance. 
The solution is obtained as follows:
\begin{equation}
\hat{\bar{\mathbf{w}}}=\frac{R^{-1}\bar{\mathbf{v}}}{\bar{\mathbf{v}}^{\HT}R^{-1}\bar{\mathbf{v}}},
\end{equation}
where $\bar{\mathbf{v}}=[\mathbf{v}^{\top},0,0,\ldots,0]^{\top}$ is a column vector containing $\mathbf{v}$ followed by $M(L_w-b+1)$ zeros, and $R$ is a power-normalized temporal-spatial covariance matrix with a prediction delay, which is defined as
\begin{equation}
R=\sum_{t}\frac{\bar{\mathbf{x}}_{t}\bar{\mathbf{x}}_{t}^{\HT}}{\sigma_{t}^{2}}.
\end{equation}
Finally, with the estimated convolutional filter, $\hat{\bar{\mathbf{w}}}$, the target speech is estimated as 
\begin{equation}
\hat{d}_{t}^{(q)}=\hat{\bar{\mathbf{w}}}^{\HT}\bar{\mathbf{x}}_{t}.
\end{equation}

Interestingly, the same solution can be derived for the proposed method even when we concatenate {\MPDR} and {\WPE} in reverse order. The signal obtained in this case becomes
\begin{equation}
\hat{d}_{t}^{(q)}=\mathbf{w}_{0}^{\HT}\mathbf{x}_{t}-\sum_{\tau=b}^{L_w}\mathbf{c}_{\tau}^{\HT}(W_{0}^{\HT}\mathbf{x}_{t-\tau}),
\end{equation}
where $\mathbf{w}_{0}$ is the MPDR beamformer applied to $\mathbf{x}_{t}$, $W_{0}$ is an arbitrary denoising matrix that contains $\mathbf{w}_{0}$ in its first column, and $\mathbf{c}_{t}$ is a coefficient vector that predicts the current denoised signal, $\mathbf{w}_{0}^{\HT}\mathbf{x}_{t}$, from the past denoised signals, $W_{0}^{\HT}\mathbf{x}_{t-\tau}$. Then, eq.~(\ref{eq:composite}) is obtained by setting $\mathbf{w}_{t}=-W_{0}\mathbf{c}_{t}$, and optimized in the way discussed above. 



\begin{figure}[!t]
\centering
\includegraphics[width=3in]{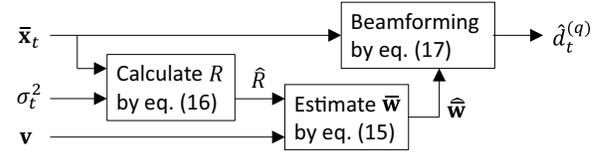}
\caption{Processing flow of {\WPD} (proposed method).}
\label{fig_sim}
\end{figure}

\section{Experiments}

\subsection{Dataset and evaluation metrics}
%
We evaluated the performance of the proposed method using the REVERB Challenge dataset \cite{REVERB}. The evaluation set (Eval set) of the dataset is composed of simulated data (SimData) and real recordings (RealData). Each utterance in the dataset contains reverberant speech uttered by a speaker and stationary additive noise. The distance between the speaker and the microphone array is varied from 0.5~m to 2.5~m. For SimData, the reverberation time is varied from about 0.25 s to 0.7 s, and the signal-to-noise ratio (SNR) is set at about 20 dB.

As objective measures for evaluating speech enhancement performance \cite{metrics}, we used the cepstrum distance (CD), 
the frequency-weighted segmental SNR (FWSSNR), the speech-to-reverberation modulation energy ratio (SRMR) \cite{SRMR}, and the speech intelligibility in bits with the information capacity of a Gaussian channel (SIIB$^{\mbox{\tiny Gauss}}$) \cite{siib}.  SIIB$^{\mbox{\tiny Gauss}}$ is a recently proposed intrusive instrumental metric that is used to evaluate the intelligibility of distorted speech signals.
To evaluate the enhanced speech in terms of ASR performance, we used a baseline ASR system recently developed using kaldi \cite{kaldi}. This is a fairly competitive system composed of a time-delay neural network acoustic model trained using a lattice-free maximum mutual information criterion and online  i-vector extraction, and a tri-gram language model.

\subsection{Methods to be compared and analysis conditions}
We compared {\WPD} (Proposed) with {\WPE}, {\MPDR}, and {\WPE} followed by {\MPDR} (WPE+MPDR). For all the methods, a hanning window was used for a short time analysis with the frame length and the shift set at 32~ms and 8~ms, respectively. The sampling frequency was 16 kHz and $M=8$ microphones were used. For {\WPE}, WPE+MPDR, and {\WPD}, the prediction delay was set at $b=4$, and the order of the autoregressive model was set at $L_w=12, 10, 8$, and $6$, respectively, for frequency ranges of $0$ to $0.8$ kHz, $0.8$ to $1.5$ kHz, $1.5$ to $3$ kHz, $3$ to $6$ kHz, and $6$ to $8$ kHz.

The time-varying power, $\sigma_{t}^2$, and the steering vector, $\mathbf{v}$ were estimated from the captured signal based on a method used in \cite{lukas2018interspeech}.  Figure~\ref{fig_sim} shows the processing flow. The same estimates were used for all the methods. Adopting the power of the captured signal as the initial value of $\sigma_{t}^2$, we repeatedly applied WPE+MPDR to the captured signal, and updated $\mathbf{v}$ and $\sigma_{t}^2$ using the outputs of the {\WPE} and {\MPDR}, respectively. The number of iterations was set at two. The steering vector was estimated based on the generalized eigenvalue decomposition with covariance whitening \cite{ito17icassp,Gannot15icassp} assuming that each utterance has noise-only periods of 225 ms and 75 ms, respectively, at its beginning and ending parts. 

\begin{figure}[!t]
\centering
\includegraphics[width=2.5in]{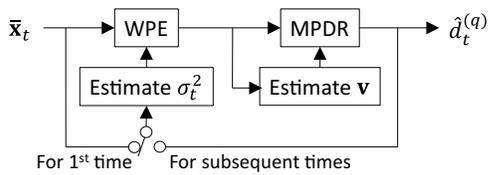}
\caption{Processing flow for estimating $\sigma_{t}^2$ and $\mathbf{v}$ by iterating WPE+MPDR.}
\label{fig_sim}
\end{figure}

\subsection{Evaluation with objective  speech enhancement measures}

Table~\ref{tbl:aq} summarizes evaluation results obtained using objective speech enhancement measures. First, all the methods improved the speech quality with all the measures. In addition, WPE+MPDR greatly outperformed {\WPE} and {\MPDR}, while the proposed method further outperformed WPE+MPDR for all the metrics except for SRMR on SimData. These results clearly show the superiority of {\WPD}.

\begin{table}[!t]
\renewcommand{\arraystretch}{1.2}
\caption{Objective quality of enhanced speech evaluated using REVERB Challenge eval set. No Enh means no speech enhancement. Boldface indicates the best score for each metric. }\label{tbl:aq}
\label{table_example}
\centering
\begin{tabular}{|c|cccc|c|}
\hline
  & \multicolumn{4}{c|}{SimData} & RealData\\
\cline{2-6}
       & CD & SRMR&  FWSSNR & SIIB$^{\mbox{\tiny Gauss}}$ & SRMR\\\hline
No Enh   & 3.97 & 3.68 &  3.62 & 241.2 & 3.18 \\
WPE      & 3.76 & 4.77 &  4.99 & 315.3 & 5.00 \\
MPDR     & 3.67 & 4.50 &  4.66 & 312.4 & 4.82 \\
WPE+MPDR & 3.01 & {\bf 5.37} & 7.52 & 486.8 & 6.57 \\
Proposed & {\bf 2.64} & 5.34 &  {\bf 8.18} & {\bf 521.7} & {\bf 6.64} \\
\hline
\end{tabular}
\end{table}

\subsection{Evaluation using ASR}

Table~\ref{tbl:wer} shows the word error rates (WERs) obtained using the baseline ASR system. 
The proposed method greatly outperformed all the other methods under all the conditions.  

\begin{table}[!t]
\renewcommand{\arraystretch}{1.3}
\caption{Word error rate (WER) in \% evaluated using REVERB Challenge eval set. No Enh means no speech enhancement. Boldface indicates the best score for each condition. }\label{tbl:wer}
\centering
\renewcommand{\arraystretch}{1.2}
\begin{tabular}{|c|cc|c|cc|c|}
\hline
  & \multicolumn{3}{c|}{SimData} &\multicolumn{3}{c|}{RealData}
\\\cline{2-7}
 & Near & Far & Average & Near & Far & Average \\\hline
No Enh & 4.18 & 6.25 & 5.22 & 17.53 & 19.68 & 18.61\\
WPE & 4.04 & 4.90 & 4.47 & 12.33 & 13.88 & 13.11 \\
MPDR & 3.81 & 4.65 & 4.23 & 10.60 & 13.81 & 12.20 \\
WPE+MPDR & 4.00 & 4.69 & 4.35 & 8.75 & 11.31 & 10.03 \\
Proposed & {\bf 3.60} & {\bf 3.95} & {\bf 3.78} & {\bf 7.86} & {\bf 10.67} & {\bf 9.27}\\
\hline
\end{tabular}

\end{table}

Finally, it may be interesting to compare {\WPD} roughly\footnote{The analysis conditions used for the two methods, such as the length of the convolutional filter and the way of calculating $\sigma_{t}^2$ and $\mathbf{v}$, are not the same.} with the frontend of the best performing system \cite{delcroix15eurasip} at the REVERB challenge.  The frontend was composed of {\WPE} and {\MVDR} followed by a nonlinear denoising method, DOLPHIN \cite{nakatani13taslp}.  With this frontend and the kaldi ASR baseline, the average WERs for RealData were 10.29 and 9.07 \%  w/o and w/ DOLPHIN, respectively. In contrast, when we evaluated {\WPD} w/o and w/ DOLPHIN, the WERs were 9.27 and 8.91~\%, respectively. This again indicates the superiority of {\WPD}.

\section{Concluding remarks}
This paper presented a method for unifying {\WPE} and {\MPDR} that made it possible to perform denoising and dereverberation both optimally and simultaneously based on microphone array signal processing. Convolutional beamforming by WPD was derived and shown to improve the speech enhancement performance in noisy reverberant environments, with regard to objective speech enhancement measures and WERs, in comparison with conventional methods, including {\WPE}, {\MPDR}, and WPE+MPDR. Future work will include an evaluation of {\WPD} in various environments, the introduction of different optimization criteria, and the extension of the proposed method to online processing.


%

\bibliographystyle{IEEEbib}
\bibliography{bibs}

\begin{thebibliography}{10}

\bibitem{Anguera07ASLP}
X.~Anguera, C.~Wooters, and J.~Hernando,
\newblock ``Acoustic beamforming for speaker diarization of meetings,''
\newblock {\em IEEE Trans. ASLP}, vol. 15, no. 7, pp. 2011--2022, 2007.

\bibitem{MPDR}
H.~L.~V. Trees,
\newblock {\em Optimum Array Processing, Part IV of Detection, Estimation, and
  Modulation Theory},
\newblock Wiley-Interscience, New York, 2002.

\bibitem{Cox}
H.~Cox,
\newblock ``Resolving power and sensitivity to mismatch of optimum array
  processors,''
\newblock {\em The Journal of the Acoustical Society of America}, vol. 54, pp.
  771--785, 1973.

\bibitem{cgmmmvdr}
T.~Higuchi, N.~Ito, S.~Araki, T.~Yoshioka, M.~Delcroix, and T.~Nakatani,
\newblock ``Online {MVDR} beamformer based on complex {Gaussian} mixture model
  with spatial prior for noise robust {ASR},''
\newblock {\em IEEE/ACM Transactions on Audio, Speech, and Language
  Processing}, vol. 25, no. 4, pp. 780--793, 2017.

\bibitem{erdogan16intersp}
H.~Erdogan, J.~R. Hershey, S.~Watanabe, M.~I. Mandel, and J.~Le Roux,
\newblock ``Improved {MVDR} beamforming using single-channel mask prediction
  networks,''
\newblock {\em Proc. Interspeech}, pp. 1981--1985, 2016.

\bibitem{L1MPDR}
S.~Emura, S.~Araki, T.~Nakatani, and N.~Harada,
\newblock ``Distortionless beamforming optimized with $l_1$-norm
  minimization,''
\newblock {\em IEEE Signal Processing Letters}, vol. 25, no. 7, pp. 936--940,
  2018.

\bibitem{GEV}
E.~Warsitz and R.~Haeb-Umbach,
\newblock ``Blind acoustic beamforming based on generalized eigenvalue
  decomposition,''
\newblock {\em IEEE Transactions on Audio, Speech, and Language Processing},
  vol. 15, no. 5, 2007.

\bibitem{maxSNR}
S.~Araki, H.~Sawada, and S.~Makino,
\newblock ``Blind speech separation in a meeting situation with maximum {SNR}
  beamformer,''
\newblock {\em Proc. IEEE ICASSP}, pp. 41--44, 2007.

\bibitem{BeamNET}
J.~Heymann, L.~Drude, C.~Boeddeker, P.~Hanebrink, and R.~Haeb-Umbach,
\newblock ``Beamnet: end-to-end training of a beamformer-supported multichannel
  {ASR} system,''
\newblock {\em Proc. IEEE ICASSP}, pp. 5235--5329, 2017.

\bibitem{wpe}
T.~Nakatani, T.~Yoshioka, K.~Kinoshita, M.~Miyoshi, and B.-H. Juang,
\newblock ``Speech dereverberation based on variance-normalized delayed linear
  prediction,''
\newblock {\em IEEE Transactions on Audio, Speech, and Language Processing},
  vol. 18, no. 7, pp. 1717--1731, 2010.

\bibitem{gwpe}
T.~Yoshioka and T.~Nakatani,
\newblock ``Generalization of multi-channel linear prediction methods for blind
  {MIMO} impulse response shortening,''
\newblock {\em IEEE Transactions on Audio, Speech and Language Processing},
  vol. 20, no. 10, pp. 2707--2720, 2012.

\bibitem{owpe}
T.~Yoshioka, H.~Tachibana, T.~Nakatani, and M.~Miyoshi,
\newblock ``Adaptive dereverberation of speech signals with speaker-position
  change detection,''
\newblock {\em Proc. IEEE ICASSP}, pp. 3733--3736, 2009.

\bibitem{swpe}
A.~Juki\'{c}, T.~van Waterschoot, T.~Gerkmann, and S.~Doclo,
\newblock ``Multi-channel linear prediction-based speech dereverberation with
  sparse priors,''
\newblock {\em IEEE/ACM Transactions on Audio, Speech and Language Processing},
  vol. 23, no. 9, pp. 1509--1520, 2015.

\bibitem{cwpe}
D.~Giacobello and T.~L. Jensen,
\newblock ``Speech dereverberation based on convex optimization algorithms for
  group sparse linear prediction,''
\newblock {\em Proc. IEEE ICASSP}, pp. 446--450, 2018.

\bibitem{REVERB}
K.~Kinoshita, M.~Delcroix, S.~Gannot, E.~A.~P. Habets, R.~Haeb-Umbach,
  W.~Kellermann, V.~Leutnant, R.~Maas, T.~Nakatani, B.~Raj, A.~Sehr, and
  T.~Yoshioka,
\newblock ``A summary of the {REVERB} challenge: state-of-the-art and remaining
  challenges in reverberant speech processing research,''
\newblock {\em EURASIP Journal on Advances in Signal Processing}, vol.
  doi:10.1186/s13634-016-0306-6, 2016.

\bibitem{CHiME3}
J.~Barker, R.~Marxer, E.~Vincent, and S.~Watanabe,
\newblock ``The third {`CHiME'} speech separation and recognition challenge:
  {D}ataset, task and baselines,''
\newblock {\em Proc. IEEE ASRU-2015}, pp. 504--511, 2015.

\bibitem{CHiME4}
E.~Vincent, S.~Watanabe, J.~Barker, and R.~Marxer,
\newblock ``{CHiME4 Challenge},''
  \url{http://spandh.dcs.shef.ac.uk/chime_challenge/chime2016/}.

\bibitem{CHiME5}
J.~Barker, S.~Watanabe, and E.~Vincent,
\newblock ``{CHiME5 Challenge},''
  \url{http://spandh.dcs.shef.ac.uk/chime_challenge/}.

\bibitem{GoogleHome}
B.~Li, T.~N. Sainath, J.~Caroselli, A.~Narayanan, M.~Bacchiani, A.~Misra,
  I.~Shafran, H.~Sak, G.~Pundak, K.~Chin, K.~Sim, R.~J. Weiss, K.~W. Wilson,
  E.~Variani, C.~Kim, O.~Siohan, M.~Weintraub, E.~McDermott, R.~Rose, and
  M.~Shannon,
\newblock ``Acoustic modeling for {Google} {Home},''
\newblock {\em Proc. Interspeech}, 2017.

\bibitem{homepod}
Audio Software Engineering {and} Siri~Speech Team,
\newblock ``Optimizing {Siri} on {HomePod} in far-field settings,''
\newblock {\em Apple Machine Learning Journal}, vol. 1, no. 12, 2018.

\bibitem{SPM}
R.~Haeb-Umbach, S.~Watanabe, T.~Nakatani, M.~Bacchiani, B.~Hoffmeister,
  M.~Seltzer, and M.~Souden,
\newblock ``Speech processing for digital home assistants,''
\newblock {\em submitted to IEEE Signal Processing Magazine}, 2019.

\bibitem{delcroix15eurasip}
M.~Delcroix, T.~Yoshioka, A.~Ogawa, Y.~Kubo, M.~Fujimoto, N.~Ito, K.~Kinoshita,
  M.~Espi, S.~Araki, T.~Hori, and T.~Nakatani,
\newblock ``Strategies for distant speech recognition in reverberant
  environments,''
\newblock {\em EURASIP J. Adv. Signal Process}, vol. Article ID 2015:60,
  doi:10.1186/s13634-015-0245-7, 2015.

\bibitem{mvdrwpe}
W.~Yang, G.~Huang, W.~Zhang, J.~Chen, and J.~Benesty,
\newblock ``Dereverberation with differential microphone arrays and the
  weighted-prediction-error method,''
\newblock {\em Proc. IWAENC}, 2018.

\bibitem{Togami}
M.~Togami,
\newblock ``Multichannel online speech dereverberation under noisy
  environments,''
\newblock {\em Proc. EUSIPCO}, pp. 1078--1082, 2015.

\bibitem{lukas2018interspeech}
L.~Drude, C.~Boeddeker, J.~Heymann, R.~Haeb-Umbach, K.~Kinoshita, M.~Delcroix,
  and T.~Nakatani,
\newblock ``Integrating neural network based beamforming and weighted
  prediction error dereverberation,''
\newblock {\em Proc. Interspeech}, pp. pp.~3043--3047, 2018.

\bibitem{wpegsc}
T.~Dietzen, S.~Doclo, M.~Moonen, and T.~van Waterschoot,
\newblock ``Joint multi-microphone speech dereverberation and noise reduction
  using integrated sidelobe cancellation and linear prediction,''
\newblock {\em Proc. IWAENC}, 2018.

\bibitem{ito17icassp}
N.~Ito, S.~Araki, M.~Delcroix, and T.~Nakatani,
\newblock ``Probabilistic spatial dictionary based online adaptive beamforming
  for meeting recognition in noisy and reverberant environments,''
\newblock {\em Proc. IEEE ICASSP}, pp. 681--685, 2017.

\bibitem{Gannot15icassp}
S.~Markovich-Golan and S.~Gannot,
\newblock ``Performance analysis of the covariance subtraction method for
  relative transfer function estimation and comparison to the covariance
  whitening method,''
\newblock pp. 544--548, 2015.

\bibitem{Nak-ICASSP2008}
T.~Nakatani, T.~Yoshioka, K.~Kinoshita, M.~Miyoshi, and Juang B.H.,
\newblock ``Blind speech dereverberation with multi-channel linear prediction
  based on short time {Fourier} transform representation,''
\newblock {\em Proc. IEEE ICASSP}, pp. 85--88, 2008.

\bibitem{early}
J.~S. Bradley, H.~Sato, and M.~Picard,
\newblock ``On the importance of early reflections for speech in rooms,''
\newblock {\em The Journal of the Acoustic Sociaty of America}, vol. 113, pp.
  3233--3244, 2003.

\bibitem{pe97}
K.~Abed-Meraim and P.~Loubaton,
\newblock ``Prediction error method for second-order blind identification,''
\newblock {\em IEEE Trans. on Signal Processing}, vol. 45, no. 3, pp. 694--705,
  1997.

\bibitem{mstep}
K.~Kinoshita, M.~Delcroix, T.~Nakatani, and M.~Miyoshi,
\newblock ``Suppression of late reverberation effect on speech signal using
  long-term multiple-step linear prediction,''
\newblock {\em IEEE Transactions on Audio, Speech, and Language Processing},
  vol. 17, no. 4, pp. 534--545, 2009.

\bibitem{RTF}
I.~Cohen,
\newblock ``Relative transfer function identification using speech signals,''
\newblock {\em IEEE Trans. on Speech, and Audio Processing}, vol. 12, no. 5,
  pp. 451--459, 2004.

\bibitem{MPDRREVERB}
J.~Heymann, L.~Drude, and R.~Haeb-Umbach,
\newblock ``A generic neural acoustic beamforming architecture for robust
  multi-channel speech processing,''
\newblock {\em Computer, Speech, and Language}, vol. 46, pp. 374--385, 2017.

\bibitem{metrics}
Y.~Hu and P.~C. Loizou,
\newblock ``Evaluation of objective quality measures for speech enhancement,''
\newblock {\em IEEE T-ASLP}, vol. 16, no. 1, pp. 229--238, 2008.

\bibitem{SRMR}
T.~H. Falk, C.~Zheng, and W.~Y. Chan,
\newblock ``A non-intrusive quality and intelligibility measure of reverberant
  and dereverberated speech,''
\newblock {\em IEEE T-ASLP}, vol. 18, no. 7, pp. 1766--1774, 2010.

\bibitem{siib}
S.~Van Kuyk, W.~B. Kleijn, and R.~C. Hendriks,
\newblock ``An evaluation of intrusive instrumental intelligibility metrics,''
\newblock {\em IEEE/ACM Trans. on Audio, Speech, and Language Processing},
  2019.

\bibitem{kaldi}
D.~Povey, A.~Ghoshal, G.~Boulianne, L.~Burget, O.~Glembek, N.~Goel,
  M.~Hannemann, P.~Motlicek, Y.~Qian, P.~Schwarz, J.~Silovsky, G.~Stemmer, and
  K.~Vesely,
\newblock ``The kaldi speech recognition toolkit,''
\newblock {\em Proc. IEEE ASRU}, 2011.

\bibitem{nakatani13taslp}
T.~Nakatani, S.~Araki, T.~Yoshioka, M.~Delcroix, and M.~Fujimoto,
\newblock ``Dominance based integration of spatial and spectral features for
  speech enhancement,''
\newblock {\em IEEE Trans. ASLP}, vol. 21, no. 12, pp. 2516--2531, 2013.

\end{thebibliography}

\end{document}